\documentclass{article}
\usepackage{graphicx}
\usepackage{sidecap}
\usepackage{subfigure}
\usepackage{amssymb}
\usepackage{amsthm}
\usepackage{amsmath}
\usepackage{bm}             
\usepackage[mathscr]{eucal}
\usepackage{cancel}
\usepackage{psfrag}
\usepackage{wasysym}

\def\pmb#1{\setbox0=\hbox{$#1$}%
  \kern-.025em\copy0\kern-\wd0
  \kern.05em\copy0\kern-\wd0
  \kern-.025em\raise.0433em\box0}
\def\pmbs#1{\setbox0=\hbox{$\scriptstyle #1$}%
  \kern-.0175em\copy0\kern-\wd0
  \kern.035em\copy0\kern-\wd0
  \kern-.0175em\raise.0303em\box0}
\def\be{\begin{equation}}
\def\ee{\end{equation}}
\def\bea{\begin{eqnarray}}
\def\eea{\end{eqnarray}}
\def\lb{\label}

\def\bi{\bibitem}
\def\vec#1{\mbox{\boldmath$#1$}}

\def\gam{\gamma}
\def\d{\delta}
\def\eps{\epsilon}

\def\sig{\sigma}

\def\Sig{\Sigma}

\def\Om{\Omega}

\def\vece{\vec{e}}

\def\la{\langle}
\def\ra{\rangle}

\def\hsp5{\hspace{5mm}}
\def\case#1/#2{\textstyle\frac{#1}{#2}}
\newcommand{\sfrac}[2]{{\textstyle{#1\over#2}}}

\theoremstyle{plain}
\newtheorem{theorem}{Theorem}[section]

\newtheorem{proposition}[theorem]{Proposition}
\newtheorem{conjecture}[theorem]{Conjecture}

\theoremstyle{remark}

\title{\sc Tilted two-fluid Bianchi type I models}
\author{\sc
 Patrik Sandin $^{1}$\thanks{Electronic address: {\tt patrik.sandin@kau.se}}\\
$^{1}${\small\em Department of Physics, University of Karlstad,}\\
{\small\em S-651 88 Karlstad, Sweden}}

\begin{document}
\maketitle

\begin{abstract}

In this paper we investigate expanding Bianchi type I models
with two tilted fluids with the same linear equation of state,
characterized by the equation of state parameter $w$.
Individually the fluids have non-zero energy fluxes w.r.t. the
symmetry surfaces, but these cancel each other because of the
Codazzi constraint. We prove that when $w=0$ the model
isotropizes to the future. Using numerical simulations and a
linear analysis we also find the asymptotic states of models
with $w>0$. We find that future isotropization occurs if and
only if $w\leq \sfrac{1}{3}$. The results are compared to
similar models investigated previously where the two fluids
have different equation of state parameters.

\end{abstract}

\centerline{\bigskip\noindent PACS numbers: 04.20.-q, 04.20.Dw,
04.20.Ha, 98.80.-k, 98.80.Bp, 98.80.Jk} \vfill
\newpage

\section{Introduction}\label{Sec:intro}

There have been numerous investigations of spatially
homogeneous (SH) cosmological models with a perfect fluid as a
matter source. In the study of these models one usually
distinguish between two different scenarios: when the fluid
congruence is normal to the hypersurfaces of homogeneity and
when it is not. The models of the first kind are called
\emph{orthogonal} or \emph{non-tilted}, and the second kind are
called \emph{tilted} models. The orthogonal models have been
studied extensively over the past few decades (see \cite{WE}
and references therein), while the tilted have received
detailed investigations only more
recently~\cite{hewittetal}-\cite{colher02}. In investigations
of tilted Bianchi models, the models of type I have been
ignored since they are not compatible with any net energy flux
with respect to the homogeneous hypersurfaces, something a
tilted fluid will induce. However, it is possible to consider
models of type I with tilted fluids if one have two or more
fluids; the fluids can then all be tilted and induce energy
flux individually, but the total energy flux must vanish. This
was done for two tilted perfect fluids with linear equations of
state in \cite{sandinuggla}. The fluids had equations of state
$p_{(1)}=w_{(1)}\,\rho_{(1)}$ and
$p_{(2)}=w_{(2)}\,\rho_{(2)}$, where $\rho_{(i)}$ and $p_{(i)}$
are the energy densities and pressures of the respective fluids
in their own rest frames, and where it was assumed that $0 \leq
w_{(2)} < w_{(1)} < 1$. The case $w_{(1)} = w_{(2)}$ was left
due to lack of space and because of its qualitatively different
behavior -- this will be the topic of the present paper.

As an example of a model with two different perfect fluids with
the same equation of state one can imagine a universe filled
with two kinds of dust (the pressure of both fluids is zero).
This is a reasonable model for both the baryonic and cold dark
matter components of the total matter content of the universe
today according to observations. The study of Bianchi type I
models with two tilted fluids with the same equation of state
can be viewed as the simplest anisotropic tilted model
generalizing the standard isotropic cosmological scenario close
to flatness.

The SH Bianchi models admit a three dimensional group of
isometries acting simply transitively on the spacelike
hypersurfaces that define the surfaces of homogeneity. The
models are classified by the Lie algebra of the isometry group,
which results in a hierarchy of models of different complexity
where the Bianchi type I model can be found on the bottom, its
algebra being Abelian and obtainable from all the other Bianchi
models by Lie algebra contractions \cite{WE}. When the Einstein
field equations are formulated as a dynamical system the
Bianchi type I models appear as a boundary of a larger state
space describing the other Bianchi models. Describing the
dynamics of this boundary is an important first step to
understanding the dynamics of the more general models.

The Bianchi models are sometimes interesting even when studying
general inhomogeneous cosmological models. This is the case for
example in the very early universe near an initial singularity
where horizons form and asymptotically shrink towards the
singularity which result in the equations asymptotically
approaching those of homogenous models, something referred to
as \emph{asymptotic silence} and \emph{locality}
\cite{pastattractor}.

In this paper we analyze the system of autonomous DEs derived
in \cite{sandinuggla} describing the time evolution of the
spatially homogeneous Bianchi type I models with two perfect
fluids, but in contrast to \cite{sandinuggla} we study the case
when the two equations of state are the same. The paper is
organized as follows: In section 2 the system of equations and
constraints is derived. In section 3 the future evolution of
dust models is described, and in section 4 the evolution of
models with other equation of state is investigated. Section 5
consists of a summary of the results and a discussion of their
implication. Appendix A contains a description of the
equilibrium points of the system and their stability
properties, and appendix B studies the dynamics on the vacuum
subset.

\section{The dynamical system}
\label{Sec:dynsysr}

We begin by deriving the evolution equations for the tilted
two-fluid Bianchi type I models. We first give the evolution
equations of a general Bianchi model, in terms of
expansion-normalized variables defined relative to the timelike
congruence normal to the group orbits. This derivation is
described in \cite{WE}, or \cite{HvECG}, but for completeness
we give a short description of it here.

First one introduces a group-invariant orthonormal frame
$\{\mathbf{e}_0, \, \mathbf{e}_{\alpha}\}$, where $\mathbf{e}_0
= \mathbf{n}$ is the normal to the group orbits and
$\alpha=1,2,3$. The commutation functions, $\gamma^a{}_{bc}(t)$
are the basic variables:

\begin{displaymath}\label{A0}
[\mathbf{e}_b,\,\mathbf{e}_c] = \gamma^a{}_{bc}\mathbf{e}_a.
\end{displaymath}

The commutator functions are normally decomposed according to

\begin{subequations}
\begin{align} \lb{dcomts0} [\,\vece_{0}, \vece_{\alpha}\,] &= -
[\,H\,\d_{\alpha}{}^{\beta} + \sig_{\alpha}{}^{\beta} +
\eps_{\alpha}{}^{\beta}{}_{\gam}\,\Omega^{\gam}\,]\,
\vece_{\beta}\:, \\
\lb{dcomtsa} [\,\vece_{\alpha}, \vece_{\beta}\,] &=
c^\gam{}_{\alpha\beta}\,\vece_{\gam} =
2a_{[\alpha}\,\d_{\beta]}{}^{\gam} +
\eps_{\alpha\beta\delta}\,n^{\delta\gam}\:.
\end{align}
\end{subequations}
where $H$ is the Hubble scalar, which is related to the
expansion $\theta$ of the normal congruence $\mathbf{n}$
according to $H=\frac{1}{3}\theta$; $\sigma_{\alpha\beta}$ is
the shear associated with ${\bf n}$; $\Omega^\alpha$ is the
Fermi rotation which describes how the spatial triad rotates
with respect to a gyroscopically fixed so-called Fermi frame;
$n^{\alpha\beta}$ and $a_\alpha$ describe the Lie algebra of
the 3-dimensional simply transitive Lie group and determine the
spatial three-curvature, see e.g.~\cite{WE}.

The energy-momentum tensor is similarly decomposed,

\begin{equation}\label{A2}
T_{ab} = \rho\,n_a\,n_b + 2q_{(a}\,n_{b)} + p\,(g_{ab} +
n_a\,n_b) + \pi_{ab},
\end{equation}

\noindent and is described by the source terms relative to the
orthonormal frame,

\begin{displaymath}\label{A3}
\{\rho, \, p, \, q_{\alpha}, \, \pi_{\alpha\beta}\}\:.
\end{displaymath}

To obtain a regular, dimensionless system of equations the
commutation functions and the source terms are normalized with
the hubble scalar according to

\begin{eqnarray}\label{A4}
\Sigma_{\alpha\beta} &=& \frac{\sigma_{\alpha\beta}}{H}\:, \quad N_{\alpha\beta} =
\frac{n_{\alpha\beta}}{H}\:, \quad A_{\alpha} = \frac{a_{\alpha}}{H}\:, \quad R_{\alpha}
= \frac{\Omega_{\alpha}}{H}\:, \nonumber \\
\Omega &=& \frac{\rho}{3H^2}\:, \quad P = \frac{p}{3H^2}\:, \quad Q_{\alpha} =
\frac{q_{\alpha}}{3H^2}\:, \quad \Pi_{\alpha\beta} = \frac{\pi_{\alpha\beta}}{3H^2}\:.
\end{eqnarray}
We also choose a new dimensionless time coordinate $\tau$
according to

\begin{equation}\label{A5}
\frac{d\tau}{dt} = H.
\end{equation}

\noindent The evolution of $H$ is determined by the
deceleration parameter $q$,

\begin{equation}\label{A6}
H^\prime = -(1+q)H\:,
\end{equation}

\noindent where $~{}^\prime$ denotes differentiation with
respect to $\tau$. Raychaudhuri's equation gives an expression
for $q$ in terms of the variables (\ref{A4}):

\begin{equation}\label{decelpar}
q = 2\Sigma^2 + \sfrac{1}{2}(\Omega + 3P),
\end{equation}

\noindent where
$\Sigma^2=\sfrac{1}{6}\Sigma_{\alpha\beta}\,\Sigma^{\alpha\beta}$.
The Einstein field equations and the Jacobi identities then
yields the following system of equations:

\vspace*{2mm} \noindent {\em Evolution equations}:
\begin{subequations}\lb{devoleq}
\begin{align}
\Sigma_{\alpha\beta}^\prime &= -(2-q)\Sigma_{\alpha\beta} +
2\epsilon^{\gamma\delta}{}_{\la \alpha}\,\Sigma_{\beta\ra
\delta}\,R_\gamma - \,^3{\cal R}_{\la\alpha\beta\ra} +
3\Pi_{\alpha\beta}\:,
\lb{HspatE}\\
A_{\alpha}^\prime &= [q\,\d_{\alpha}{}^{\beta} -
\Sig_{\alpha}{}^{\beta} - \eps_{\alpha}{}^{\beta}{}_{\gam}\,
R^{\gam}] A_\beta\:,\lb{Hajac}\\
(N^{\alpha\beta})^\prime &= [q\,\d_{\gamma}{}^{(\alpha} +
2\Sig_{\gam}{}^{(\alpha} + 2
\eps_{\gam}{}^{(\alpha}{}_{\delta}\,R^{\delta}] N^{\beta )
\gamma}\lb{Hnjac}\:.
\end{align}
\end{subequations}

\vspace*{2mm} \noindent {\em Constraint equations}:
\begin{subequations}\lb{dconstreqsil}
\begin{align}
0 &= 1 - \Sigma^2 + \sfrac{1}{6}\,^3{\cal R} - \Omega\:,\label{dGauss}\\
0 &= (3\delta_\alpha{}^\gamma\,A_\beta +
\epsilon_\alpha{}^{\delta\gamma}
\,N_{\delta\beta})\,\Sigma^\beta{}_\gamma - 3Q_\alpha\:,\label{dCodazzi}\\
0 &= A_\beta\, N^\beta{}_\alpha\:.\lb{HJacobi}
\end{align}
\end{subequations}

\noindent where $^3{\cal R}_{\la\alpha\beta\ra}$ and $^3{\cal
R}$ are the trace-free and scalar parts of the
Hubble-normalized three-curvature, respectively, according to:

\begin{displaymath}\lb{threecurv} ^3{\cal R}_{\la\alpha\beta\ra} = B_{\la
\alpha\beta \ra} + 2\epsilon^{\gamma\delta}{}_{\la
\alpha}\,N_{\beta\ra\delta}\,A_\gamma\:,\quad ^3{\cal R} =
-\sfrac{1}{2}B^\alpha{}_\alpha - 6A_{\alpha}A^{\alpha}\:;\qquad
B_{\alpha\beta} = 2 N_{\alpha\gamma}\,N^\gamma{}_\beta -
N^\gamma{}_\gamma\,N_{\alpha\beta}\:,
\end{displaymath}
and where $\la..\ra$ denotes trace-free symmetrization of the
indices, i.e. $A_{\la\alpha\beta\ra} =
A_{(\alpha\beta)}-\sfrac{1}{3}\delta_{\alpha\beta}A_{\alpha}A^{\beta}$.
The contracted Bianchi identities yields evolution equations
for the total source variables:

\begin{subequations}\lb{dmattereq}
\begin{align}
\lb{dlomdot} \Om^\prime &=  (2q-1)\,\Om - 3P +
2A_{\alpha}\,Q^{\alpha} - \Sig_{\alpha\beta}\Pi^{\alpha\beta}\:,\\
\lb{dlqmalpha} Q_{\alpha}^\prime &= -[2(1-q)\,\d_{\alpha}{}^{\beta}
+ \Sig_{\alpha}{}^{\beta} +
\eps_{\alpha}{}^{\beta}{}_{\gam}\,R^{\gam}]\,Q_{\beta} +
(3\delta_\alpha{}^\gamma\,A_\beta +
\epsilon_\alpha{}^{\delta\gamma}
\,N_{\delta\beta})\,\Pi^\beta{}_\gamma\:.
\end{align}
\end{subequations}

\noindent These equations are a consequence of the evolution
and constraint equations \eqref{devoleq} and
\eqref{dconstreqsil} and give no additional information, but
are useful as auxiliary equations.

The system of equations \eqref{devoleq} and
\eqref{dconstreqsil} with \eqref{decelpar} is not fully
determined, there are no evolution equations for the variables
$R_{\alpha}$ that represent the angular velocity of the spatial
frame $\{\mathbf{e}_{\alpha}\}$. We can freely specify
$R_{\alpha}$ in any way most convenient. Also the evolution of
the source variables $P$ and $\Pi_{\alpha\beta}$ are not
determined until we specify a source. We consider the source to
be two tilted perfect fluids, only interacting with each other
gravitationally. We can then split the energy-momentum tensor
into two parts, each satisfying the conservation equation
separately

\begin{equation}
T^{ab} = \sum_{i}T_{(i)}^{ab}\:, \qquad \nabla_a T_{(i)}^{ab}=0\:,
 \qquad (i=1,\,2)\:,
\end{equation}

\noindent where

\begin{equation}
T^{ab}_{(i)} = (\tilde{\rho}_{(i)} + \tilde{p}_{(i)})
\tilde{u}^a_{(i)}\tilde{u}^b_{(i)} + \tilde{p}_{(i)} g^{ab}\:.
\end{equation}

\noindent We now impose the same linear equation of state for
the two fluids, i.e., $\tilde{p}_{(i)} = w \tilde{\rho}_{(i)}$,
where $w=const$. The four velocities $\mathbf{u}_{(i)}$ can be
written in the form

\begin{equation}
\tilde{u}^a_{(i)} = \frac{1}{\sqrt{1 - v_{a(i)}v^a_{(i)}}}(n^a + v^a_{(i)})\:;
\qquad n_a v^a_{(i)}=0\:.
\end{equation}
Since we have two separate conservation equations we now have
two sets of source variables

\begin{displaymath}
\{\Omega_{(i)}, \, P_{(i)}, \, Q^{(i)}_{\alpha}, \,
\Pi^{(i)}_{\alpha} \}, \qquad (i=1,\,2)\:,
\end{displaymath}

\noindent where each $\Omega_{(i)}$ and $Q^{(i)}_{\alpha}$
satisfies equations (\ref{dmattereq}). The source variables can
all be expressed in terms of the normalized energy density and
the three-velocity of the respective fluid as

\begin{subequations}\label{matter}
\begin{align}
Q_{(i)}^\alpha &= (1 + w)(G^{(i)}_+)^{-1}\,
v_{(i)}^\alpha\,\Omega_{(i)}\:,\\
P_{(i)} &= w\Omega_{(i)} + \sfrac{1}{3} (1 - 3w)
Q^{(i)}_\alpha v_{(i)}^\alpha\:,\\
\Pi^{(i)}_{\alpha\beta} &= Q^{(i)}_{\la\alpha}v^{(i)}_{\beta\ra}\:,
\end{align}
\end{subequations}

\noindent where $G^{(i)}_\pm  =  1 \pm w \, v_{(i)}^2$ and
$v^2_{(i)} = v_{\alpha(i)}v^{\alpha}_{(i)}$. One can use
equations (\ref{dmattereq}) and (\ref{matter}) to obtain
evolution equations for the source terms $\Omega_{(i)}$ and
$v^{\alpha}_{(i)}$:
\begin{subequations}
\begin{align}
\Omega_{(i)}^\prime &= (2q - 1 - 3w)\,\Omega_{(i)} + [(3w-1)\,
v_{(i)\alpha} -
\Sigma_{\alpha\beta}\,v_{(i)}^\beta + 2A_\alpha]\,Q_{(i)}^\alpha\:,\\
v^\prime_{\alpha(i)} &= (G^{(i)}_-)^{-1}\,\left[ (1-v_{(i)}^2)
(3w - 1 - w\,A^\beta\,v_{\beta(i)}) + (1-w)(A^\beta +
\Sigma_\gamma{}^\beta\,v_{(i)}^\gamma)\,v_{\beta(i)} \right]
v_{\alpha(i)} \nonumber \\
& \quad - [\Sigma_\alpha{}^\beta +
\epsilon_\alpha{}^{\beta\gamma}\,(R_\gamma +
N_\gamma{}^\delta\,v_{\delta(i)})]\,v_{\beta(i)} - A_\alpha\,v_{(i)}^2
\:.\lb{pecsil}
\end{align}
\end{subequations}

For the Bianchi type I models we have $A_{\alpha} =
N_{\alpha\beta} = 0$. The Codazzi constraint (\ref{dCodazzi})
then becomes, $Q_\alpha= Q_\alpha^{(1)} + Q_\alpha^{(2)}= 0$,
which, taken in combination with~\eqref{matter}, forces the
3-velocities of the two fluids to be anti-parallel.
Kinematically the situation is similar to that of Bianchi type
I with a general magnetic field studied in~\cite{leb97}, and it
is therefore natural to exploit the same mathematical
structures in the present problem. We therefore choose the
spatial triad so that one of the frame vectors is aligned with
the fluid velocities, which we choose to be $\vece_3$, i.e.
$v_{(i)}^{\alpha}= (0,0,v_{(i)})$. Demanding that these
conditions on $v_{(i)}^{\alpha}$ hold for all times lead to the
following conditions
\be R_1 = -\Sigma_{23}\:,\qquad R_2 = \Sigma_{31}\:. \ee
This leaves $R_3$ undetermined, however, we still have the
freedom of arbitrary rotations in the 1-2-plane, which we use
to set
\be R_3=0\:.\ee
Following~\cite{leb97}, we introduce the variables
$\Sigma_+,\Sigma_A,\Sigma_B,\Sigma_C$ according to
\be \Sigma_+ = \sfrac{1}{2}(\Sigma_{11}+\Sigma_{22})\:,\qquad
\Sigma_{31} + \mathrm{i}\,\Sigma_{23} =  \sqrt{3}\Sigma_A\,
e^{\mathrm{i}\phi} \:,\qquad  \Sigma_-  +
\frac{\mathrm{i}}{\sqrt{3}} \Sigma_{12}= (\Sigma_B +
\mathrm{i}\,\Sigma_C) e^{2\mathrm{i}\phi}\:, \ee
where $\Sigma_-=(\Sigma_{11}-\Sigma_{22})/(2\sqrt{3})$, which
leads to
\be \Sigma^2= \Sigma_+^2 + \Sigma_A^2 + \Sigma_B^2 +
\Sigma_C^2\:. \ee
The angular variable $\phi$ decouples from the other equations,
$\phi^\prime = -\Sigma_C$, which reduces the dimension of the
dynamical system by one. The resulting dynamical system is the
following:

\vspace*{2mm} \noindent {\em Evolution equations}:
\begin{subequations} \label{evolBI}
\begin{align}
\Sigma_+^\prime &= -(2-q)\Sigma_+ + 3\Sigma_A^2 -  Q_{(1)}
v_{(1)} -  Q_{(2)}v_{(2)}\:, \label{Sigp}\\
\Sigma_A^\prime &= -(2 - q + 3\Sigma_+ + \sqrt{3}\Sigma_B)
\Sigma_A \:, \label{SigA}\\
\Sigma_B^\prime &= -(2-q)\Sigma_B + \sqrt{3}\Sigma_A^2 -
2\sqrt{3}\Sigma_C^2\:, \\
\Sigma_C^\prime &= -(2 - q - 2\sqrt{3}\Sigma_B) \Sigma_C\:,
\label{SigC}\\
v_{(i)}^\prime &= (G^{(i)}_-)^{-1} (1-v_{(i)}^2)(3w_{(i)} - 1 +
2\Sigma_+)v_{(i)}\:, \label{vieq}\\
\Omega_{(i)}^\prime &= (2q - 1 - 3w_{(i)})\Omega_{(i)} +
(3w_{(i)} - 1 + 2\Sigma_+)Q_{(i)}v_{(i)}\:.
\label{Omieq}
\end{align}
\end{subequations}

\vspace*{2mm} \noindent {\em Constraint equations}:
\begin{subequations} \label{constrBI}
\begin{align}
0 &= 1 -\Sigma^2 - \Omega_{(1)} - \Omega_{(2)}
\:,\label{gausssys}\\
0 &= Q_{(1)} + Q_{(2)}\label{codazzisys}\:,
\end{align}
\end{subequations}
where
\be\lb{qeq} q = 2\Sigma^2 + \sfrac{1}{2}(\Omega_{\rm m}  +
3P_{\rm m}) = 2 - \sfrac{3}{2}(\Omega_{\rm m} - P_{\rm m})\:;
\quad \Omega_{\rm m} = \Omega_{(1)} + \Omega_{(2)}\:, \quad
P_{\rm m} = P_{(1)} + P_{(2)}\:. \ee
The assumption of non-negative energy densities, $\Omega_{(i)}
\geq 0$, together with~\eqref{qeq} and~\eqref{gausssys}, yields
that $\sfrac{1}{2} \leq q \leq 2$.

\subsection{The state space}\label{statespace}

The state space consists of $\mathbf{S} =
\{\Sigma_+,\,\Sigma_A,\,\Sigma_B,\,\Sigma_C,\,v_{(1)},\,v_{(2)},\,
\Omega_{(1)},\,\Omega_{(2)}\}$ subject to the constraints
(\ref{gausssys}), (\ref{codazzisys}). Both $v_{(i)}=0$ and
$v_{(i)}^2=1$ defines invariant subsets which means that
$v^2_{(i)}$, contained in the range $(0,1)$, is bounded both
from above and below. The constraint (\ref{gausssys}) together
with the non-negativity condition on the energy densities puts
bounds on all the other variables and hence the state space is
bounded. We are primarily interested in the interior state
space where

\be 0 < \Omega_{(1)}\Omega_{(2)}, \quad 0 < v_{(i)}^2 < 1, \ee
but asymptotically the orbits of the system may approach the
boundary and hence we consider its closure, $\bar{{\bf S}}$,
which means that we consider the set defined by $\Sigma^2 \leq
1, 0\leq v_{(i)}^2\leq 1$; $0\leq\Omega_{(i)}\leq 1$, in such a
way that the constraints~\eqref{constrBI} are satisfied.

The dynamical system have the following discrete symmetries:
\be\label{discrete} \Sigma_A \rightarrow -\Sigma_A\:,\quad
\Sigma_C \rightarrow -\Sigma_C\:;\qquad (v_{(1)},v_{(2)})
\rightarrow -(v_{(1)},v_{(2)})\:. \ee
We therefore assume without loss of generality that $\Sigma_A
\in [0,1], \, \Sigma_C \in [0,1], \, v_{(1)} \in [0,1]$, and
$v_{(2)} \in [-1,0]$; the solutions in the other sectors of the
state space are easily obtained by means of the discrete
symmetries.

\begin{figure}[h]
\psfrag{a}[ll][ll]{$0$}
\psfrag{b}[ll][ll]{$k \to \ldots $}
\psfrag{c}[ll][ll]{}
\psfrag{d}[ll][ll]{}
\psfrag{e}[ll][ll]{$1$}
\psfrag{f}[ll][ll]{}
\psfrag{g}[ll][ll]{}
\psfrag{h}[ll][ll]{$\infty$...}
\psfrag{A}[cc][cc]{$v_{(1)}$}
\psfrag{B}[cc][cc]{$v_{(2)}$}
\centering
        \subfigure[Dust, $w=0$.]{
        \includegraphics[height=0.40\textwidth]{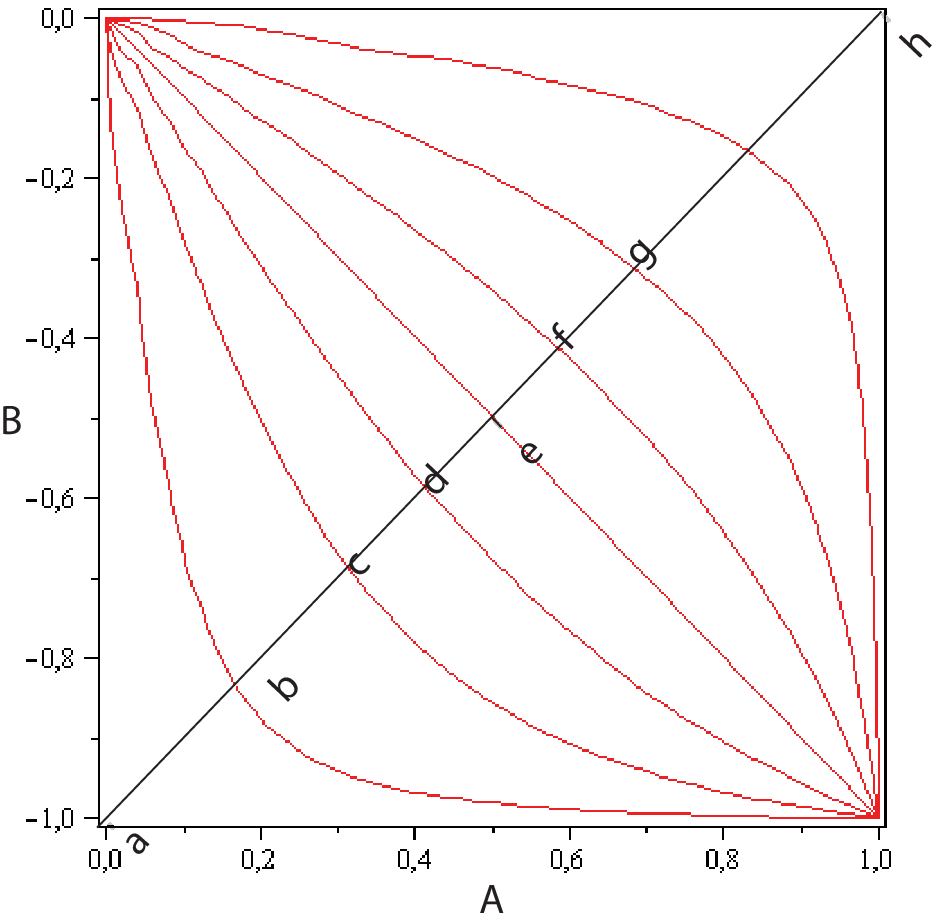}}\qquad
        \subfigure[Almost stiff, $w = 0.99$.]{
        \includegraphics[height=0.40\textwidth]{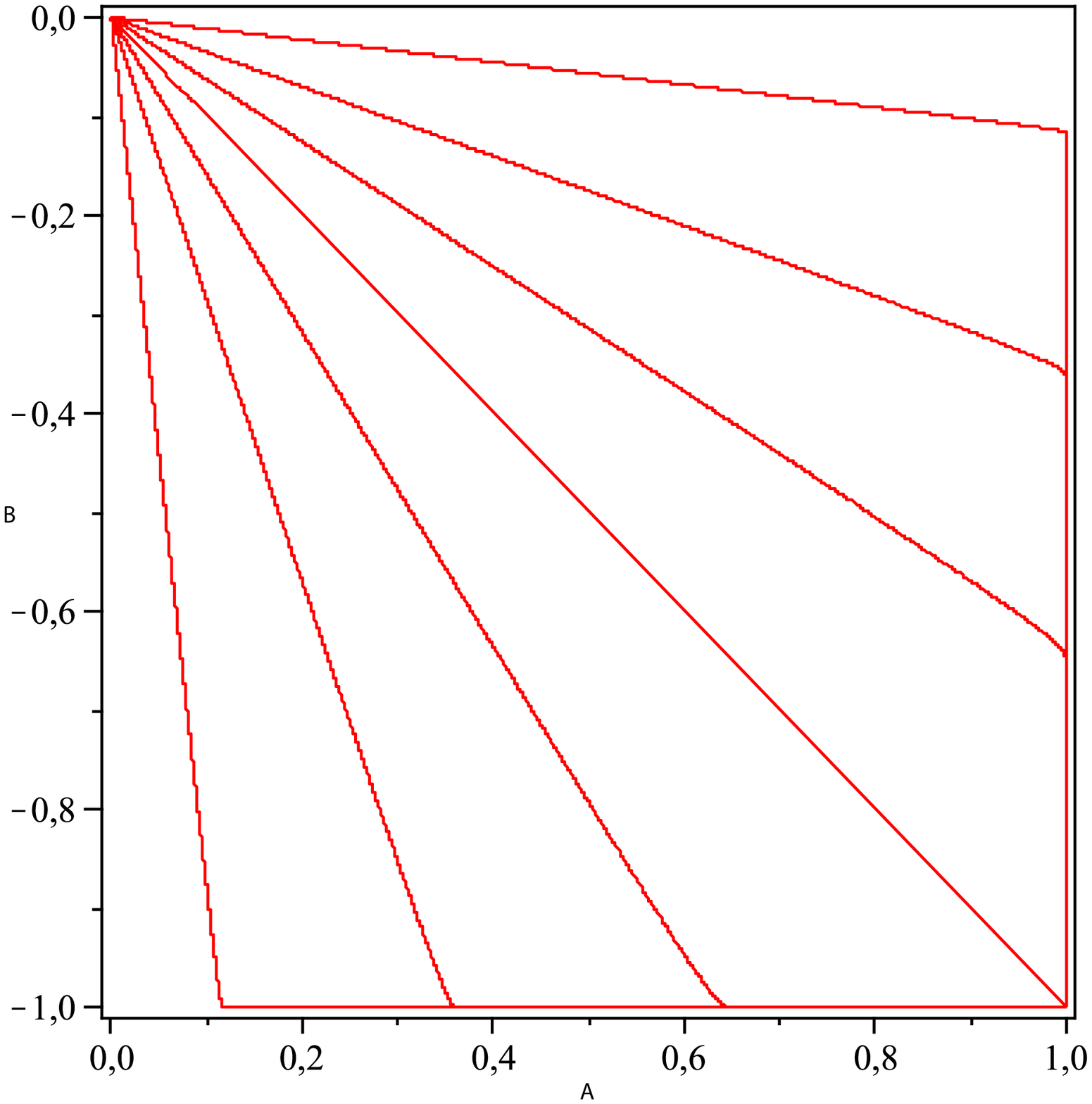}}\qquad
        \caption{The leaves of foliation in $v_{(1)}
        \times v_{(2)}$-space.}
        \label{Figure1}
\end{figure}

In \cite{sandinuggla} one had two monotone functions which here
become constants of motion. Only one of them is independent of
the Codazzi constraint, however, and can be written as
\be \label{phi} \frac{v_{(1)}^2(1-v_{(2)}^2)^{(1-w)}}{v_{(2)}^2
(1-v_{(1)}^2)^{(1-w)}}= k. \ee
where $k$ is a positive real constant. The submanifolds defined
by (\ref{phi}) foliates the state space and are dependent on
$w$. Projections onto $v_{(1)} \times v_{(2)}$-space for dust
and almost stiff equations of state are shown in Figure
\ref{Figure1}.

The constant of motion (\ref{phi}) correlates the velocities of
the fluids such that if one of them becomes extremely tilted or
orthogonal asymptotically so must the other. Note also that it
is not possible that one of the fluids will dominate over the
other asymptotically; if the normalized energy density of one
of them vanishes then this is also the case for the other,
which follows from considering the constraints (\ref{phi}) and
(\ref{codazzisys}) simultaneously. This excludes a large part
of the boundary of $\bar{{\bf S}}$ from consideration since it
may not be approached. When examining invariant subsets and fix
points on the boundary in the following we will only consider
those that can be approached from the interior, i.e. not those
where $\Omega_{(i)}=0 \neq \Omega_{(j)}$, $v_{(i)}=0 \neq
v_{(j)}$, or $v_{(i)}^2=1 \neq v_{(j)}^2$, where $i \neq j$.

\subsection{Invariant subsets}

The dynamical system~\eqref{evolBI},~\eqref{constrBI}, admits a
number of invariant subsets, conveniently divided into two
classes: (i) `geometric subsets', i.e., sets associated with
conditions on the shear and hence the metric; (ii) invariant
sets associated with conditions on the tilt or energy
densities. We will introduce a notation where the kernel
suggests the type of subset and where a subscript, when
existent, suggests the values of $v_{(1)}$ and $v_{(2)}$.

\hspace{1mm}

{\em Geometric subsets}
\begin{itemize}
\item ${\cal TW}$: The `twisting' subset, characterized by
    $\Sigma_C=0,\, \Sigma_{A}\neq 0$, which leads to that
    the decoupled $\phi$-variable satisfies $\phi=const$
    and hence $\Sigma_{12} \propto \Sigma_{11}-\Sigma_{22}$.
\item ${\cal RD}$: The constantly rotated diagonal subset,
    given by $\Sigma_A=0,\,\Sigma_C\neq 0$ ($R_\alpha=0$).
    This subset is the diagonal subset, discussed next,
    rotated with a constant angle around $\vece_3$.
\item ${\cal D}$: The diagonal subset, defined by
    $\Sigma_A=\Sigma_C=0;\, \Sigma_B = \Sigma_-$, and hence
    $R_\alpha=0$.
\item ${\cal LRS}$: The locally rotationally symmetric
    subset. This plane symmetric subset of the diagonal
    subset is characterized by the additional condition
    $\Sigma_B=\Sigma_-=0$. This is the simplest subset
    compatible with two tilted fluids.
\end{itemize}

\hspace{1mm}

{\em Matter subsets}
\begin{itemize}
\item ${\cal O}$: The orthogonal subset for which
    $v_{(1)}=v_{(2)}=0$. In this subset $\Omega_{(1)}
    \propto \Omega_{(2)}$ and the distinction between the
    two fluids becomes artificial. The orthogonal subset
    describes a single orthogonal fluid. In general this
    subset is expressed in a non-Fermi frame for which
    $\Sigma_A\Sigma_C\neq 0$, however, usually when dealing
    with this case one makes a rotation to a Fermi frame in
    which the shear and the metric are diagonal so that
    ${\cal O}$ belongs to ${\cal D}$.
\item ${\cal ET}_{11}$: The double extreme tilt subset
    where both fluids propagate with the speed of light,
    $v_{(1)}=1=-v_{(2)}\,\Rightarrow\, \Omega_{(1)} =
    \Omega_{(2)}=3P_{(1)}=3P_{(2)}$.
\item ${\cal K}$: The vacuum subset is called the Kasner
    subset and is defined by $\Omega_{\rm m}=0;
    \,\Sigma^2=1$; it describes the Kasner solutions, but
    in general in a non-Fermi propagated frame, and with
    $v_{(i)}$ as test fields.
\end{itemize}
Other invariant subsets can be obtained by taking intersections
of the ones described above.

Apart from the above subsets there are also a number of fix
points. These, and their eigenvalues, are given in
Appendix~\ref{locstab}, but we summarize them in
Table~\ref{fixpoints} along with their stability properties. If
$\Omega_{(1)}$ and $\Omega_{(2)}$ are zero then $v_{(1)}$ and
$v_{(2)}$ act as test fields. This is the case for the Kasner
fix points ${\rm K}_{**}$ which all describe the same Kasner
solutions; the ${\rm KS}_{v_{(1)v_{(2)}}}^\pm$ describe a
particular Kasner solution determined by $w$; the line of
Friedmann fix points ${\rm F}_{00}^{\Omega_{(1)}\Omega_{(2)}}$
describe the flat Friedmann-Lema\^itre solution for a single
orthogonal perfect fluid with a linear equation of state.
However, all the remaining fix points are physically distinct.

\begin{table}
\renewcommand{\arraystretch}{1.45}
\begin{tabular}{|l|c|c|c|c|c|}
\hline
Name & $0\leq\ w < \sfrac{1}{3}$ & $\sfrac{1}{3}
< w < \sfrac{1}{2}$ & $\sfrac{1}{2}
< w < \sfrac{5}{9}$ & $\sfrac{5}{9}$ & $\sfrac{5}{9}
< w < 1$\\
\hline
${\rm K}^{\ocircle}_{00}$ & \multicolumn{5}{c|}{source/saddle}\\
${\rm K}^{\ocircle}_{11}$ & \multicolumn{5}{c|}{source/saddle}\\
\hline
${\rm KS}_{v_{(1)}v_{(2)}}^\pm$ & \multicolumn{5}{c|}
{center-saddle}\\
\hline
${\rm F}^{\Omega_{(1)}\Omega_{(2)}}_{00}$ & sink &
\multicolumn{4}{c|}{saddle}\\
\hline
${\rm LRSL}_{v_{(1)}v_{(2)}}$ & does not exist & sink &
\multicolumn{3}{c|}{saddle}\\
\hline
${\rm TW}_{11}$& \multicolumn{5}{c|}{saddle}\\
\hline
${\rm TWL}_{v_{(1)}v_{(2)}}$ & \multicolumn{2}{c|}
{does not exist} &  sink & $({\rm TWL}_{v_{(1)}v_{(2)}}
\subset {\rm GS}_{v_{(1)}v_{(2)}})$ & saddle\\
\cline{1-4}  \cline{6-6}
${\rm GS}_{v_{(1)}v_{(2)}}$ & \multicolumn{3}{c|}
{does not exist} & sink & does not exist\\
\cline{1-4}  \cline{6-6}
${\rm G}_{11}$ & \multicolumn{3}{c|}{saddle} & $({\rm G}_{11}
\subset {\rm GS}_{v_{(1)}v_{(2)}})$ & sink\\
\hline
\end{tabular}
\caption{A list of fix points when $0\leq w <1$. The fix points
are denoted by a kernel that is related to a subset of which
the fix point belong in combination with a subscript and
sometimes also a superscript. The subscript indicates the fix
point values of $v_{(1)}$ and $v_{(2)}$. The superscript of the
Friedmann fix points ${\rm F}_{00}^{**}$ indicates of the
values of $\Omega_{(1)}$ and $\Omega_{(2)}$ while it denotes
the sign of $\Sigma_B$ in the case of the Kasner surface ${\rm
KS}_{**}^\pm$. A complete characterization of the fix points is
given in Appendix~\ref{locstab}; here we have given a
description in terms of their stability properties.}
\label{fixpoints}
\end{table}
\section{Future isotropization of tilted multi-fluid dust models}\label{Sec:dust}

Arguably the most physically relevant equation of state is the
case when $w=0$. This describes two perfect fluids with zero
pressure, commonly referred to as \emph{dust}, and is a good
description of the actual matter content of the universe during
the matter dominated epoch. For this case we can prove the
following proposition

\begin{proposition}
If w=0, then the future asymptotic state of the system
(\ref{evolBI}) is the Friedmann-Lemaître solution.
\end{proposition}
\begin{proof}
For $w=0$ we have
\begin{equation}
[{\rm ln}\, (1-v_{(i)}^2)\Omega_{(i)}^2]^\prime = 2\,(2q-1),
\end{equation}
which is a monotonically increasing function since $q\geq
\sfrac{1}{2}$. Since it is also bounded from above it must
approach a limit value and hence we have $q \to \sfrac{1}{2}$.
This implies $P_m \to 0$, $\Omega_m \to 1$, which in turn
implies $v_{(i)} \to 0$. The constraint (\ref{gausssys})
ensures isotropization and we have reduced the system to the
fix points ${\rm FL}_{00}^{\Omega_{(1)}\Omega_{(2)}}$.
\end{proof}
Observations of galaxies and galaxy clusters correlates the
distribution of dark matter with the baryonic matter visible in
the galaxies, which implies that the velocity of dark matter is
aligned with the velocity of visible matter \cite{tyson}. The
Bianchi type I model is an example of how a universe where the
dust flows initially are non-aligned can evolve into a state
where they become aligned asymptotically to the future.

The linear analysis of appendix A, and the dynamics on the
Kasner subset, described in Appendix B and numerical
simulations suggests that asymptotically to the past the system
approaches a doubly extremely tilted Kasner model described by
one of the fix points in (\ref{Kattra}), as conjectured in
section \ref{pastattr}.

\section{Future and past dynamics of general linear equations of state}\label{Sec:attractor}

\subsection{Future dynamics}\label{futattr}

As was shown in \cite{sandinuggla} no tilted two-fluid Bianchi
type I models with $Q_{(i)}>0$, $v_{(i)}^2 < 1$ and $w >
\sfrac{1}{3}$ isotropize to the future. The theorem does not
tell us what the asymptotic state is other than that it is
anisotropic. The conclusions we make about the future global
attractors for $w < \sfrac{1}{3}$ rests on the local stability
analysis of the fix points and numerical simulations, which
makes us confident of the following conjectures:

\begin{conjecture}\label{conjecturegeneric}
The $\omega$-limit for all orbits that have $\Sigma_A,\,
\Sigma_C \neq 0$ initially is contained in the set ${\rm
FL}_{00}^{\Omega_{(1)}\Omega_{(2)}}$ if $~w \leq \frac{1}{3}$,
$~{\rm LRSL}_{v_{(1)}v_{(2)}}$ if $~\frac{1}{3} < w \leq
\frac{1}{2}$, $~{\rm TWL}_{v_{(1)}v_{(2)}}$ if $~\frac{1}{2} <
w < \frac{5}{9}$, $~{\rm GS}_{v_{(1)}v_{(2)}}$ if $~w =
\frac{5}{9}$, and $~G_{11}$ if $~\frac{5}{9} < w < 1$.
\end{conjecture}
\begin{conjecture}\label{conjecturetwist}
The $\omega$-limit for all orbits that have $\Sigma_A \neq 0$,
$\Sigma_C = 0$ initially is contained in the set ${\rm
FL}_{00}^{\Omega_{(1)}\Omega_{(2)}}$ if $~w \leq \frac{1}{3}$,
$~{\rm LRSL}_{v_{(1)}v_{(2)}}$ if $~\frac{1}{3} < w \leq
\frac{1}{2}$, $~{\rm TWL}_{v_{(1)}v_{(2)}}$ if $~\frac{1}{2} <
w < \frac{3}{5}$, and $~TW_{11}$ if $~\frac{3}{5} \leq w < 1$.
\end{conjecture}
\begin{conjecture}\label{conjecturediagonal}
The $\omega$-limit for all orbits that have $\Sigma_A = 0$
initially is contained in the set ${\rm
FL}_{00}^{\Omega_{(1)}\Omega_{(2)}}$ if $~w \leq \frac{1}{3}$
and in the set $~{\rm LRSL}_{v_{(1)}v_{(2)}}$ if $~\frac{1}{3}
< w < 1$.
\end{conjecture}
The conjectures about the future attractors can conveniently be
summarized by three diagrams showing the attractors for
different values of $w$, for the general Bianchi type I set,
the 'twisting' subset and the diagonal subset. See figure 2.
\begin{figure}[h]
\psfrag{A}[cc][cc]{$0$}
\psfrag{B}[cc][cc]{$\frac{1}{3}$}
\psfrag{C}[cc][cc]{$\frac{1}{2}$}
\psfrag{D}[cc][cc]{$\frac{5}{9}$}
\psfrag{E}[cc][cc]{$1$}
\psfrag{F}[cc][cc]{$\frac{3}{5}$}
\psfrag{a}[cc][cc]{${\rm FL}^{\Omega_{(1)}\Omega_{(2)}}_{00}$}
\psfrag{b}[cc][cc]{${\rm LRSL}_{v_{(1)}v_{(2)}}$}
\psfrag{c}[cc][cc]{${\rm TWL}_{v_{(1)}v_{(2)}}$}
\psfrag{d}[cc][cc]{${\rm GS}_{v_{(1)}v_{(2)}}$}
\psfrag{e}[cc][cc]{${\rm G}_{11}$}
\psfrag{f}[cc][cc]{$w$}
\psfrag{g}[cc][cc]{${\rm TW}_{11}$}
\psfrag{h}[cc][cc]{${\rm LRSL}_{v_{(1)}v_{(2)}}$}
\centering
        \subfigure[The general case]{
        \label{bifurLRS}
        \includegraphics[width=0.3\textwidth]{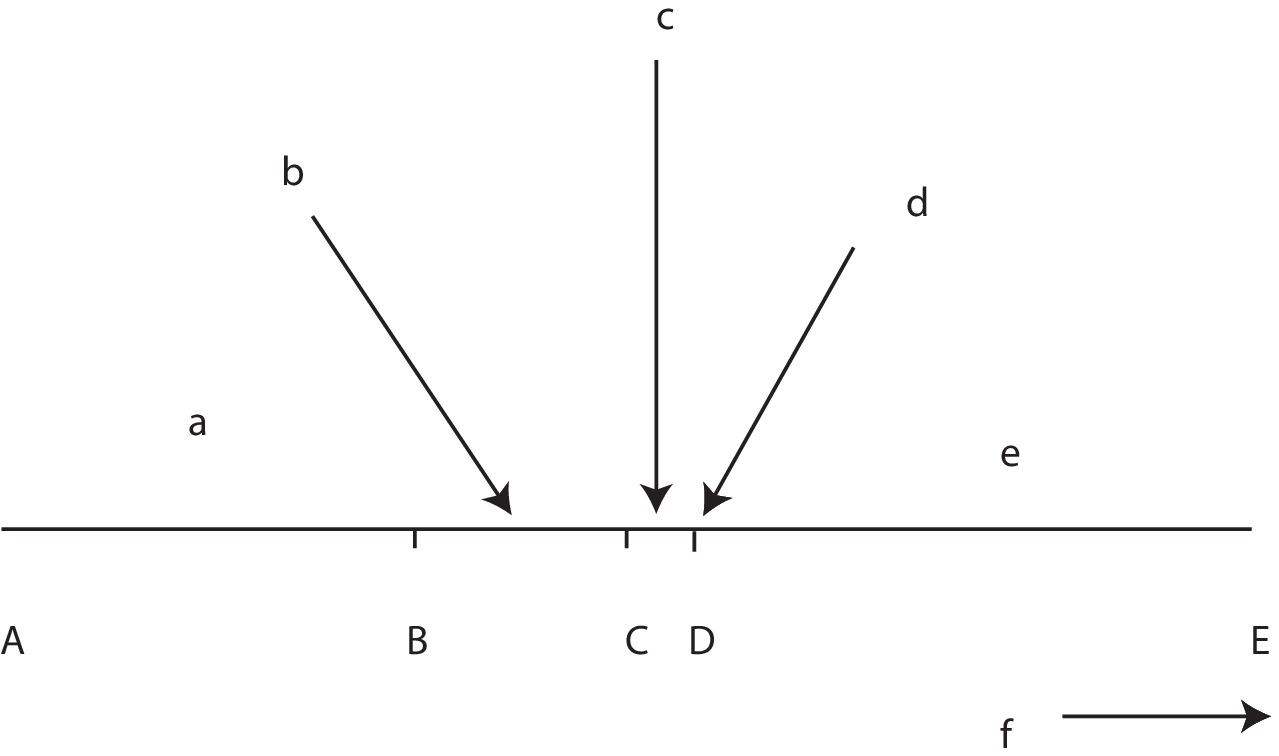}}\qquad
         \subfigure[The ${\cal TW}$ subset]{
        \label{bifurTW}
        \includegraphics[width=0.3\textwidth]{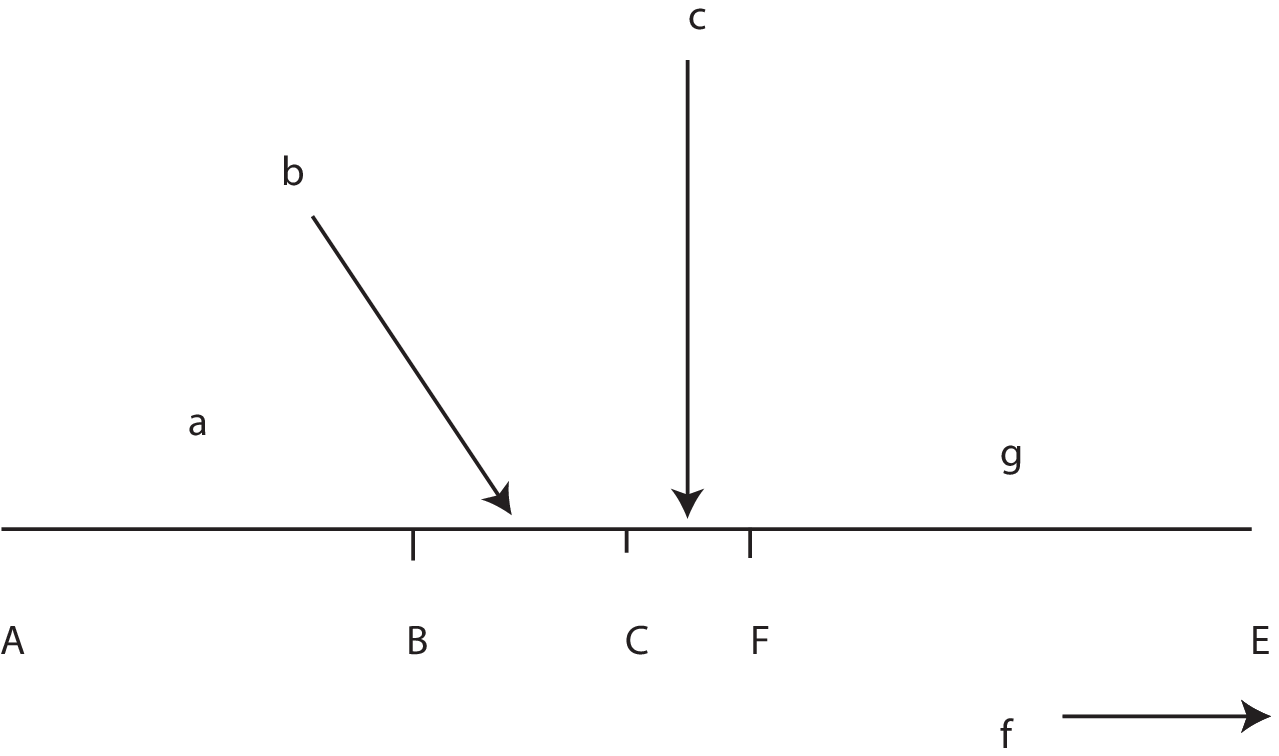}}
        \subfigure[The ${\cal RD}$, ${\cal D}$, ${\cal LRS}$ subsets]{
        \label{bifurgen}
        \includegraphics[width=0.3\textwidth]{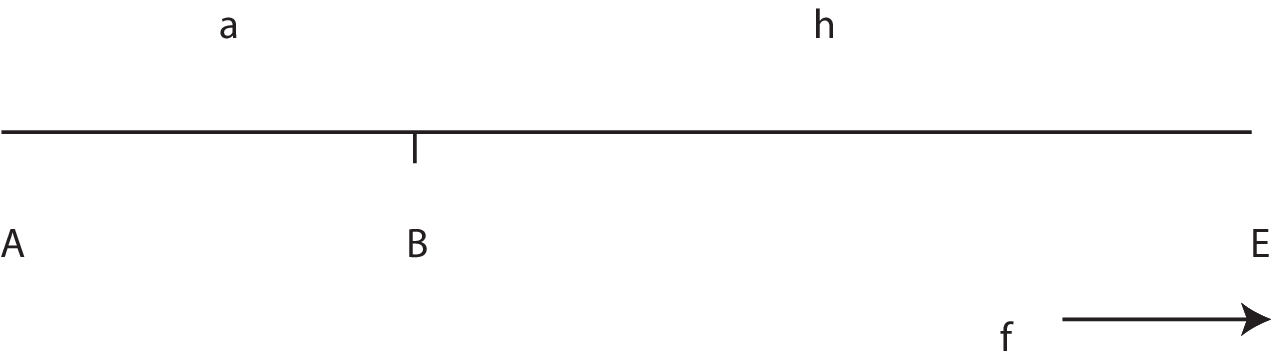}}
        \caption{Future global attractor bifurcation diagrams for the
        various geometric subsets.}
    \label{bifur}
\end{figure}
%

\subsection{Past dynamics}\label{pastattr}

Numerical calculations, the local analysis of the fix points in
appendix A, and the dynamics on the Kasner subset described in
Appendix B, supports the following conjecture:
\begin{conjecture}
The $\alpha$-limit for every orbit with $Q_{(1)}> 0,\,
v_{(1)}^2<1,\,v_{(2)}^2<1$ initially on the general geometric
set with $\Sigma_A\Sigma_C\neq 0$ is one of the fix points on
the global past attractor ${\cal A}_{\{**\}}$ for the Kasner
subset ${\cal K}$ given in equation~\eqref{Kattr}.
\end{conjecture}
All of the tilted models approach a vacuum dominated Kasner
singularity, where the influence of the matter becomes
negligible. The fluids may either become aligned with each
other and the normal congruence to the homogeneous
hypersurfaces, or anti-aligned and extremely tilted, depending
on the value of $w$ and which of the Kasner states that is
approached.

\section{Summary and discussion}\label{Sec:concl}

We have studied Bianchi type I models with two tilted fluids
with the same linear equation of state, parameterized by the
equation of state parameter $w$, using dynamical systems
methods. The absence of spatial curvature in these models
forces the fluids to be anti-aligned with each other. The paper
was written with the modest ambition of completing the analysis
of multi-fluid Bianchi type I models in \cite{sandinuggla}
where it was assumed that one the fluids were stiffer than the
other (i.e. there were two different equations of state
parameters such that $w_{(1)} > w_{(2)}$). As expected the
models described here are in many ways similar to those in
\cite{sandinuggla}, and do, for example, exhibit rather similar
bifurcation structure to the future, however , there are
differences. One important difference is the existence of a
constant of motion that correlates the velocities and energy
densities of the fluids. This prohibits the asymptotic states
to become single fluid cosmologies with an extra test field as
was the case in many situations when the two fluids had
different equations of state.

We proved that for models with $w=0$ the system isotropizes to
the future and approaches a future asymptotic Friedmann
universe where both fluids become orthogonal to the homogenous
hypersurfaces. For models with $w > 0$ we found equilibrium
points that were future stable in the full state space and
hence local future attractors. Numerical simulations indicate
that these local attractors are also global future attractors,
in the full state space and the invariant subspaces
respectively. For equations of state softer than radiation the
future attractors are the Friedmann points, corresponding to
the self similar, isotropic Friedmann universe. A bifurcation
occurs at the radiation value $w=1/3$ where the future
asymptotic state changes from a Friedmann universe to an
anisotropic, but locally rotationally symmetric, state where
both fluids acquire a non-zero tilt. There is a different fix
point for each given value of a constant of motion, described
in section \ref{statespace}. This result is different from the
case when the case where one fluid is stiffer than the other,
described in \cite{sandinuggla}, where the stiffer fluid
becomes extremely tilted. A second bifurcation occurs at
$w=1/2$ where the future asymptotic state acquires a second
component of shear. Since the future asymptotic fix point is
contained in the ´twisting´ subset, where $R_{\alpha}\neq0,\
\phi=const.$, the asymptotic state is described relative to a
constantly rotating frame. The frame is tied to the fluid
three-velocity and hence indicates that the fluid is rotating
relative to a Fermi frame. A final bifurcation occurs at
$w=5/9$ where the future asymptotic fix point is transferred
from the twisting subset to the general anisotropic state space
where $\Sigma_C \neq 0$. This implies that the decoupled
variable $\phi$ is linearly decreasing ($\phi^\prime =
-\Sigma_C$) and hence that the frame rotation vector
$R_{\alpha}$ is rotating in the plane spanned by $\mathbf{e}_1$
and $\mathbf{e}_2$. Both fluid becomes extremely tilted (i.e.
$|v_{(i)}| = 1$).

The past asymptotic state is for all non-self similar solutions
a fix point located on the vacuum subset, generically either of
zero or extreme tilt, depending on the asymptotic value of one
of the shear variables. A subset of measure zero can
asymptotically to the past have intermediate tilt if the shear
variable $\Sigma_+$ tends to a specific value.

One can ask wether the anti-alignment of the fluids is a
physically realistic scenario. Is there some mechanism which
asymptotically would produce such a state from more general
models? For past asymptotic behavior one could argue that there
might be in many cases, since Bianchi type I models seem to be
a part of the past attractor for very general cosmological
models and for these models the fluids tend to become
anti-aligned, but to really investigate wether this actually
happens one must study two-fluid models in a more general
context, one that at least contain Bianchi type II models.

\begin{appendix}

\section{Fix points and local stability
analysis}\label{locstab}
\hspace{1mm}

\noindent {\bf Kasner fix points\/}: There are two circles of
Kasner points and two surfaces of fix points when $0 \leq w$.
The Kasner circles are characterized by $\Sigma_+ =
\hat{\Sigma}_+,\, \Sigma_B = \hat{\Sigma}_-,\, \Sigma_A =
\Sigma_C = 0,\, \Omega_{(1)} = \Omega_{(2)} = 0$, where
$\hat{\Sigma}_\pm$ are constants that satisfy $\hat{\Sigma}_+^2
+ \hat{\Sigma}_-^2=1$, and the following values of $v_{(i)}$:
\begin{subequations}
\begin{align}
{\rm K}^{\ocircle}_{00}:\quad  v_{(1)} &= v_{(2)} =
0\:,\qquad\qquad\, {\rm K}^{\ocircle}_{11}:\,\,  v_{(1)} =
-v_{(2)} = 1\:.
\end{align}
\end{subequations}
The eigenvalues for the two cases are:
\begin{subequations}\label{Kasenrcirc}
\begin{align}
& {\rm K}^{\ocircle}_{00}:\quad  0\:;\quad \lambda_{\Sigma_A}\:;\quad
\lambda_{\Sigma_C}\:;\quad \lambda_{v_{(1)}}^0\:;\quad
\lambda_{v_{(2)}}^0\:;\quad 3(1-w)\:; \quad  3(1-w)\:,\\
& {\rm K}^{\ocircle}_{11}:\quad  0\:;\quad \lambda_{\Sigma_A}\:;\quad
\lambda_{\Sigma_C}\:;\quad \lambda_{v_{(1)}}^1\:;\quad
\lambda_{v_{(2)}}^1\:;\quad  2(1+\hat{\Sigma}_+)\:,
\end{align}
\end{subequations}
where
\begin{subequations}
\begin{align}
\lambda_{\Sigma_A} &=
-(3\hat{\Sigma}_+ + \sqrt{3}\hat{\Sigma}_-)\:, \qquad
\lambda_{\Sigma_C} = 2\sqrt{3}\hat{\Sigma}_-\:,\\
\lambda_{v_{(i)}}^0 &= 3w - 1 + 2\hat{\Sigma}_+\:,\qquad\,\,
\lambda_{v_{(i)}}^1 = -2(3w - 1
+ 2\hat{\Sigma}_+)/(1-w)\:.
\end{align}
\end{subequations}
In the ${\rm K}^{\ocircle}_{00}$ case the Codazzi
constraint~\eqref{codazzisys} is singular and hence it cannot
be locally solved; in the other case~\eqref{codazzisys} has
been used to eliminate $\Omega_{(1)}$. The zero eigenvalue
corresponds to that one has a one-parameter set of fixed
points. The two surfaces of Kasner fix points are characterized
by
\begin{eqnarray}\label{Kasnersurfaces}
{\rm KS}_{v_{(1)}v_{(2)}}^\pm: \quad &&\Sigma_A = \Sigma_C = \Omega_{(1)}
= \Omega_{(2)} = 0,\quad \Sigma_+ = \sfrac{1}{2}(1-3w)
\:,\quad \Sigma_B = \pm \sqrt{1-\Sigma_+^2}\:,\\ \nonumber
&&\ \left(\frac{v_{(1)}}{v_{(2)}}\right)^2 \cdot \left(\frac{1-v_{(2)}^2}
{1-v_{(1)}^2}\right)^{(1-w)}
=k \:.
\end{eqnarray}
where the superscript denotes the sign of $\Sigma_B$, and $k$
is the constant defined in~(\ref{phi}). The relation between
$v_{(1)}$, $v_{(2)}$ and $k$ constrains the three free
parameters and thus gives a surface of fix points. After
eliminating $\Omega_{(1)}$ locally by means of the Codazzi
constraint~\eqref{codazzisys}, the eigenvalues for the Kasner
surfaces are:
\begin{subequations}\label{KS}
\begin{align}
& {\rm KS}_{v_{(1)}v_{(2)}}^\pm: \quad 0\:; \quad 0\:; \quad
0\:; \quad \lambda_{\Sigma_A}\:; \quad \lambda_{\Sigma_C}\:;
\quad 3(1-w)\:,
\end{align}
\end{subequations}
where again $\lambda_{\Sigma_A} = -(3\hat{\Sigma}_+ +
\sqrt{3}\hat{\Sigma}_-)\:,\, \lambda_{\Sigma_C} =
2\sqrt{3}\hat{\Sigma}_-$, where $\hat{\Sigma}_+,\,
\hat{\Sigma}_-$ take the fix point values for the relevant line
of fix points. Here two zero eigenvalues corresponds to that
one has a surface of fix points while the third is associated
with the existence of a one parameter set of solutions that are
anti-parallel w.r.t. each other on each side of the surface of
fix points.

\hspace{1mm}

\noindent {\bf Friedmann fix points\/}: On the Friedmann subset
there exists one line of fix points parameterized  by the
constant of motion $k$~\eqref{phi}:
\begin{subequations}
\begin{align}
{\rm FL}_{00}^{\Omega_{(1)}\Omega_{(2)}}: & \quad \Sigma_+=\Sigma_A=
\Sigma_B=\Sigma_C=0, \quad \Omega_{\rm m} = 1 \quad v_{(1)}=0\:, \quad
v_{(2)}=0\:, \quad \frac{\Omega_{(1)}}{\Omega_{(2)}} = \sqrt{k} \:,
\end{align}
\end{subequations}
where the superscript refers to the values of $\Omega_{(1)}$
and $\Omega_{(2)}$. The associated eigenvalues are:
\begin{subequations}
\begin{align}
{\rm FL}_{00}^{\Omega_{(1)}\Omega_{(2)}}: & \quad \lambda_{1,2,3,4} =
-\sfrac{3}{2}(1-w)\:; \quad 3w - 1\:; \quad 0 \:,
\end{align}
\end{subequations}
were we have used the Codazzi constraint~\eqref{codazzisys} to
eliminate the variable $v_{(2)}$. Two of the eigenvalues of
$\lambda_{1,2,3,4}$ refer to $\lambda_{\Sigma_A}$ and
$\lambda_{\Sigma_C}$.

\hspace{1mm}

\noindent {\bf Fix points on ${\cal LRS}$}: When $\frac{1}{3} <
w$ there is an additional line of fix points, ${\rm
LRSL}_{v_{(1)}v_{(2)}}$, which enter the physical state space
when $w=\frac{1}{3}$, and move into the ${\cal LRS}$-subset
with increasing values of $w$. The line intersects the
foliation determined by (\ref{phi}) and can be parametrised by
$k$. We have here chosen to use $v_{(1)}$ as a parameter
instead for reasons of computational simplicity. In the stiff
perfect fluid limit ($w=1$) the line merge with the coalesced
Kasner surfaces. The line of fix points is characterized by:
\begin{subequations}
\begin{align}
& {\rm LRSL}_{v_{(1)}v_{(2)}}: \nonumber \\
& \Sigma_A = \Sigma_B =
\Sigma_C = 0\:, \quad \Sigma_+ =-\sfrac{1}{2}(3w-1)\:, \quad
v_{(2)}v_{(1)} = -\frac{3w-1}{5w+1}\:, \quad \frac{3w-1}{5w+1}
\leq v_{(1)} \leq 1\:, \nonumber \\
& \Omega_{(1)} = \frac{3}{4} \frac{(1-w)(5w+1)(3w-1)
(1+wv_{(1)}^2)}{(1+w)[(5w+1)v_{(1)}^2 + (3w-1)]}\:, \quad\
\Omega_{(2)} = \frac{3}{4}\frac{(1-w)[(5w+1)^2v_{(1)}^2 +
w(3w-1)^2]}{(1+w)[(5w+1)v_{(1)}^2 + (3w-1)]}.
\end{align}
\end{subequations}
After eliminating $\Omega_{(1)}$ locally the eigenvalues for
the LRS-line are:
\begin{subequations}
\begin{align}
{\rm LRSL}_{v_{(1)}v_{(2)}}: \quad & \lambda_{\Sigma_A}
= 3(2w-1)\:; \quad \lambda_{\Sigma_B} = \lambda_{\Sigma_C} =
-\sfrac{3}{2}(1-w)\:; \quad 0\:; \quad -\sfrac{3}{4}(1-w) \left(1 \pm
\sqrt{A(w,\ v_{(1)})} \right)\:,
\end{align}
\end{subequations}
where ${\rm Re}\, A(w,\ v_{(1)})<1$; since the expression for
$A(w_{(i)})$ is rather messy we will refrain from giving it.

\hspace{1mm}

\noindent {\bf Fix points on ${\cal TW}$\/}:

\be {\rm TW}_{11}: \,\, \Sigma_+ = -\sfrac{2}{5}\:,\,\,
\Sigma_C = 0\:, \,\, \Sigma_A = \Sigma_B =
\sfrac{\sqrt{3}}{5}\:, \quad v_{(1)}=1\:, \,\, v_{(2)} =
-1\:,\,\, \Omega_{(1)}=\Omega_{(2)}=\sfrac{3}{10}\:. \ee
Local elimination of $\Omega_{(1)}$ by means of the Codazzi
constraint~\eqref{codazzisys} yields the eigenvalues:
\be \lambda_{\Sigma_C}=\sfrac{3}{5}\:; \qquad -\sfrac{3}{5}\:;
\qquad -\sfrac{3}{10}(1 \pm i \sqrt{39})\:;\qquad
\frac{6(3-5w)}{5(1-w)}\:; \qquad \frac{6(3-5w)} {5(1-w)}\:. \ee
When $\frac{1}{2} < w < \frac{3}{5}$ there exists one more line
of fix points on ${\cal TW}$: ${\rm TWL}_{v_{(1)}v_{(2)}}$.
This line comes into existence when the line
LRSL$_{v_{(1)}v_{(2)}}$ bifurcate into two at $w =
\frac{1}{2}$; it then wanders away from ${\cal D}$ when $w$
increases and eventually leaves the physical state space
through ${\rm TW}_{11}$ when $w=\frac{3}{5}$. Like ${\rm
LRSL}_{v_{(1)}v_{(2)}}$ it also can be parameterized by $k$ but
we choose $v_{(1)}$ here also for simplicity. The fix points
are characterized by
\begin{subequations}
\begin{align}
{\rm TWL}_{v_{(1)}v_{(2)}}: \quad \Sigma_+ &=
-\sfrac{1}{2}(3w_{(1)}-1)\:,\qquad \Sigma_A =
\sqrt{\sfrac{3}{2}(1-w)(2w-1)}\:,\qquad
\Sigma_B = \sqrt{3}(2w -1)\:,\nonumber \\
\Sigma_C &= 0\:, \quad v_{(1)}v_{(2)} =
\frac{(1-w)(15w-7)}{-25w^2+18w-1}\:, \qquad
\frac{(1-w)(15w-7)}{-25w^2+18w-1} \leq v_{(1)} \leq 1\:,\nonumber\\
\Omega_{(1)} &= B(w,\,v_{(1)})\:,\qquad
\Omega_{(2)} = 1 - \sfrac{1}{4}(3w-1)(15w-7) - B(w,\,v_{(1)})\:,
\end{align}
where
\be B(w,\,v_{(1)})= \frac{3}{4}\,\frac{(1 - w)(7 - 15w) (25w^2
- 18w + 1) (1 + wv_{(1)}^2)}{(1+w)[(-25w^2 + 18w - 1)v_{(1)}^2
+ (1 - w)(7 - 15w)]}\:.\ee
\end{subequations}
Local elimination of $\Omega_{(1)}$ yields the following
eigenvalues:
\begin{subequations}
\begin{align}
& {\rm TWL}_{v_{(1)}v_{(2)}}:\,  \lambda_{\Sigma_C}=
-\sfrac{3}{2}(5-9w)\:; \quad 0\:; \quad \lambda_{3}(w,\,v_{(1)})\:;
\quad \lambda_{4}(w,\,v_{(1)})\:; \quad \lambda_{5}(w,\,v_{(1)})\:;
\quad \lambda_{6}(w,\,v_{(1)})\:,
\end{align}
\end{subequations}
where $\lambda_{3,4,5,6}$ exhibit extremely messy expressions,
which we therefore refrain from giving. They all have the
property that the real part of the eigenvalues always are
negative in the domain of definition of the fix point set, thus
making the entire line a local attractor in the range $1/2 < w
< 5/9$.

\hspace{1mm}

\noindent {\bf Fix point in the generic geometric manifold\/}:
There exists one fix point ${\rm G}_{11}$ for which all the
off-diagonal components of the shear are non-zero. It thus
exists on the generic `geometric' manifold, but on the `matter
boundary' ${\cal ET}_{11}$ where both fluids are extremely
tilted. It is characterized by:
\be {\rm G}_{11}: \,\, \Sigma_+  = -\sfrac{1}{3}\:,\,\,
\Sigma_A = \sfrac{2}{3\sqrt{3}}\:, \,\, \Sigma_B = \Sigma_C
=\sfrac{1}{3\sqrt{3}}\:,\quad v_{(1)}= 1\:,\,\, v_{(2)} =
-1\:,\,\, \Omega_{(1)} = \Omega_{(2)} = \sfrac{1}{3}\:. \ee
Local elimination of $\Omega_{(1)}$ yields the eigenvalues:
\be \lambda_{1,2,3,4} = -\sfrac{1}{3}\left(1 \pm i \sqrt{23 \pm
12\sqrt{2}}\right)\:;\qquad \lambda_{5,6} =
-\frac{2(9w-5)}{3(1-w)}\:. \ee

At $w=\frac{5}{9}$ there exists a triangular surface of fix
points, ${\rm GS}_{v_{(1)}v_{(2)}}$, connecting ${\rm
TWL}_{v_{(1)}v_{(2)}}$ with $G_{11}$. ${\rm
GS}_{v_{(1)}v_{(2)}}$ is given by:
\begin{subequations}
\begin{align}
{\rm GS}_{v_{(1)}v_{(2)}}: \quad \Sigma_+ & = -\sfrac{1}{3}\:,\qquad
\Sigma_A = \frac{\sqrt{2}}{3\sqrt{3}} \sqrt{\frac{17v_{(1)}v_{(2)}+3}
{4v_{(1)}v_{(2)}-3}}\:,\qquad \Sigma_B = \sfrac{1}{3\sqrt{3}}\:, \quad
\Sigma_C = \frac{1}{3\sqrt{3}} \sqrt{\frac{13v_{(1)}v_{(2)}+6}
{4v_{(1)}v_{(2)}-3}}\nonumber \\
-1 & \leq v_{(1)}v_{(2)} \leq -\sfrac{6}{13} \:, \nonumber \\
\Omega_{(1)} &= \frac{1}{3} \frac{-v_{(2)}(9+5v_{(1)}^2)}
{(v_{(1)}-v_{(2)})(3-4v_{(1)}v_{(2)})} \:, \quad
\Omega_{(2)} = \frac{1}{3} \frac{v_{(1)}(9+5v_{(2)}^2)}
{(v_{(1)}-v_{(2)})(3-4v_{(1)}v_{(2)})}\:.
\end{align}
\end{subequations}
Local elimination of $\Omega_{(1)}$ yields two zero eigenvalues
and four others with complicated dependence on $v_{(1)}$ and
$v_{(2)}$ but which all have negative real part on the entire
set ${\rm GS}_{v_{(1)}v_{(2)}}$.

The local stability analysis of the fix points can be
summarized in a table showing how the local attractor is
transferred from set to set with increasing values of $w$ -
from the isotropic Friedmann solutions for sub-radiation
equation of states to the increasingly anisotropic solutions
when the fluid becomes stiffer.

\hspace{1mm}

\noindent \begin{tabular}{l c c c c c c c c c}
invariant & & & & & & & & & \\
subset: & ${\cal FLO}$ & & ${\cal LRS}$ & & ${\cal TW}$ & & \multicolumn{3}{c}{GENERIC MANIFOLD} \\
local & & & & & & & & & \\
sink: & ${\rm FL}_{00}^{\Omega_{(1)}\Omega_{(2)}}$ & $\to$ &
${\rm LRSL}_{v_{(1)}v_{(2)}}$ & $\to$ &
${\rm TWL}_{v_{(1)}v_{(2)}}$ & $\to$ & ${\rm GS}_{v_{(1)}v_{(2)}}$ &
$\to$ & $G_{11}$ \\
& & & & & & & & & \\
$w$: & $\in [0,\,\sfrac{1}{3})$ & & $\in (\sfrac{1}{3}
,\,\sfrac{1}{2}]$ & & $\in (\sfrac{1}{2},\,\sfrac{5}{9})$ & &
$\sfrac{5}{9}$ & & $\in (\sfrac{5}{9},\,1)$ \\
\end{tabular}

\section{The ${\cal K}$ subset}\label{Sec:Kasner}

\begin{figure}[h]
\psfrag{a}[cc][cc]{$\Sigma_+$} \psfrag{b}[cc][cc]{$\Sigma_3$}
\psfrag{c}[cc][cc]{$\Sigma_B$} \psfrag{d}[cc][cc]{$(321)$}
\psfrag{e}[cc][cc]{$(231)$} \psfrag{f}[cc][cc]{$(213)$}
\psfrag{g}[cc][cc]{$(123)$} \psfrag{h}[cc][cc]{$(132)$}
\psfrag{i}[cc][cc]{$(312)$} \psfrag{j}[cc][cc]{0}
\psfrag{k}[cc][cc]{${\rm T}_3$}\psfrag{l}[cc][cc]{${\rm T}_2$}
\psfrag{m}[cc][cc]{${\rm T}_1$} \psfrag{n}[cc][cc]{${\rm Q}_3$}
\psfrag{o}[cc][cc]{${\rm Q}_2$}\psfrag{p}[cc][cc]{${\rm Q}_1$}
\centering
        \subfigure[Kasner sectors]{
        \label{Ksectors}
        \includegraphics[height=0.30\textwidth]{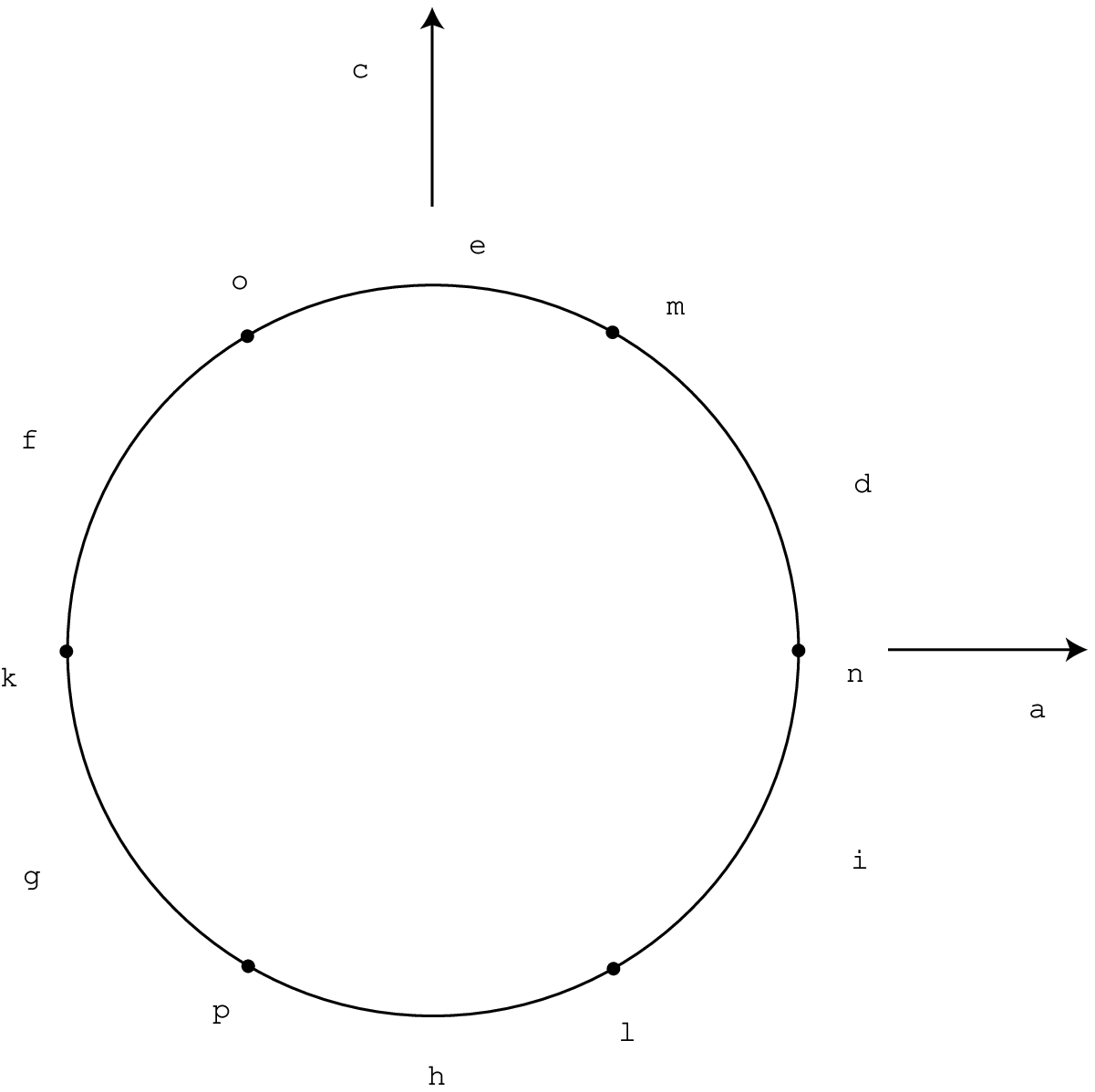}}\qquad
        \subfigure[The $\alpha$- and $\omega$-limits on ${\cal KP}$]{
        \label{Kasnersourcesink}
        \includegraphics[height=0.30\textwidth]{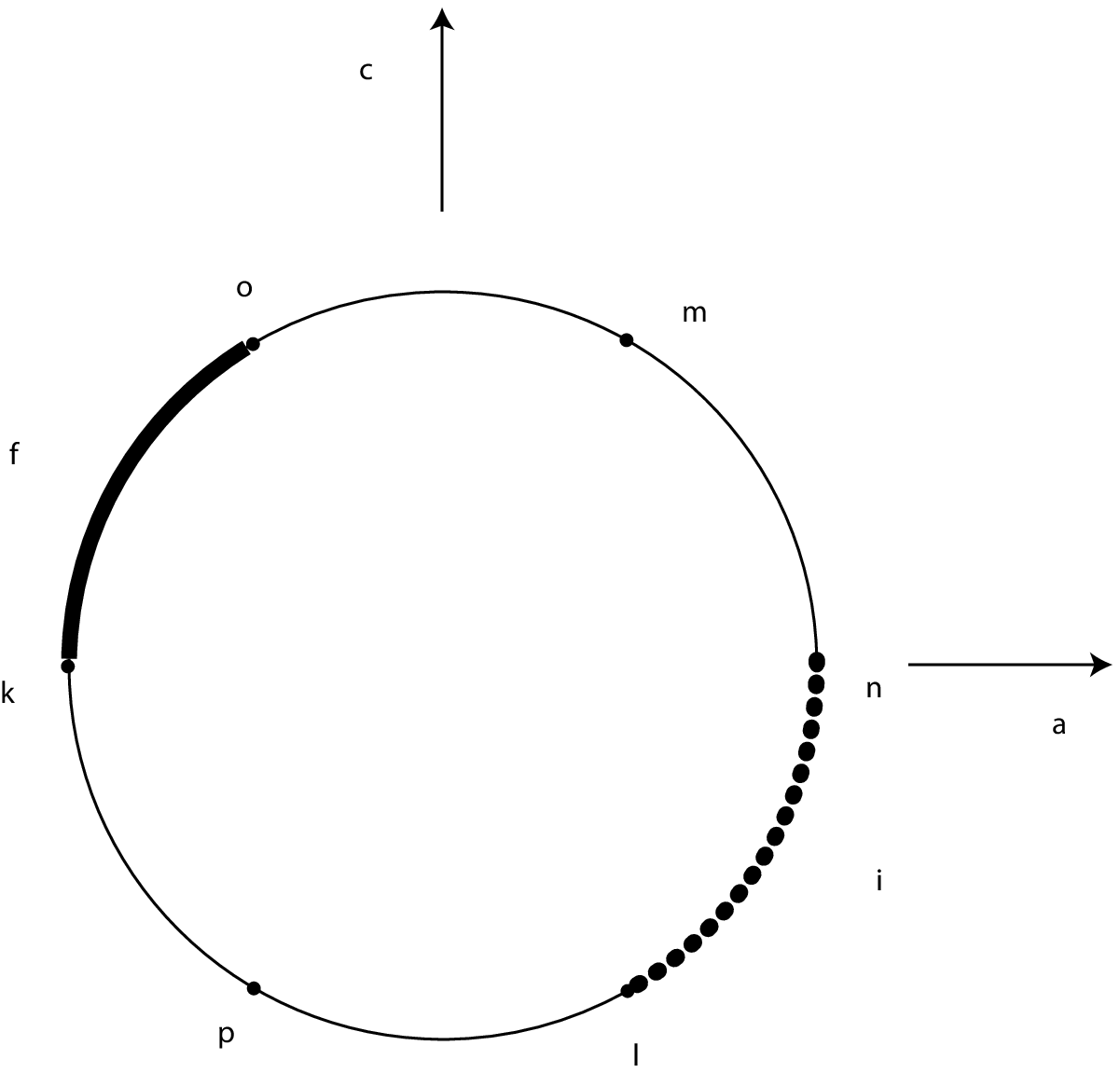}}
        \caption{The projected Kasner circle ${\rm KP}^\ocircle$ is
        divided into sectors $(i,j,k)$, defined by
        $\Sigma_i<\Sigma_j<\Sigma_k$, where $i,j,k$ is a permutation
        of $1,2,3$, and where
        $\Sigma_1=\hat{\Sigma}_+ +\sqrt{3}\hat{\Sigma}_-$,
        $\Sigma_2=\hat{\Sigma}_+ -\sqrt{3}\hat{\Sigma}_-$,
        $\Sigma_3=-2\hat{\Sigma}_+$, and the points ${\rm Q}_\alpha$,
        corresponding to the non-flat plane symmetric Kasner solution,
        and ${\rm T}_\alpha$, corresponding to the Taub form for the
        Minkowski spacetime. The global past attractor for the
        general geometric set with $\Sigma_A\Sigma_B\neq 0$ on ${\cal KP}$
        consists of sector $(213)$ together with ${\rm Q}_2$ and ${\rm T}_3$
        on ${\rm KP}^\ocircle$. The global future attractor consists
        of sector $(312)$ together with ${\rm Q}_3$ and ${\rm T}_2$
        on ${\rm KP}^\ocircle$.}
    \label{Kasner1}
\end{figure}

We here discuss the Kasner subset ${\cal K}$ with the state
space ${\bf K} =
\{\Sigma_+,\Sigma_A,\Sigma_B,\Sigma_C,v_{(1)},v_{(2)}\}$,
subjected to the constraints (\ref{gausssys}) and (\ref{phi}).
The equations for the test fields $v_{(1)} \in [0,1]$,
$v_{(2)}\in [-1,0]$ decouple from those of the shear but are
still coupled to each other through (\ref{phi}). The state
space therefore can be written as the following Cartesian
product:
\be {\bf K}= {\bf KP}\times\{v_{(1)},v_{(2)}\} \:,\qquad {\bf
KP} = \{\Sigma_+,\Sigma_A,\Sigma_B,\Sigma_C\}\:, \ee
where ${\bf KP}$ is the projected Kasner state space, which of
course is subjected to $\Sigma^2=1$. By determining the
$\alpha$- and $\omega$-limits for solutions on ${\bf KP}$ one
can then determine the asymptotic states of $v_{(1)}$ and
$v_{(2)}$, and thus the $\alpha$- and $\omega$-limits for
solutions on ${\cal K}$. Let us therefore first turn to the
equations on ${\bf KP}$:
\be\label{KP} \Sigma_+^\prime = 3\Sigma_A^2\:;\quad
\Sigma_A^\prime = -(3\Sigma_+ + \sqrt{3}\Sigma_B) \Sigma_A
\:;\quad \Sigma_B^\prime = \sqrt{3}\Sigma_A^2 -
2\sqrt{3}\Sigma_C^2\:;\quad \Sigma_C^\prime =
2\sqrt{3}\Sigma_B\Sigma_C \:. \ee
This system is defined on the compact space $\Sigma^2=1$. Since
$\Sigma_+$ is monotonically increasing on a compact space it
must approach a constant value, hence we have $ \Sigma^2_A
\propto \Sigma_+^\prime \to 0$ asymptotically both to to future
and to the past. Since we also have $(\Sigma_+ -
\sqrt{3}\Sigma_B)^\prime = -2\sqrt{3}\Sigma^2_C \leq 0$,
$\Sigma_C$ will by the same argument also vanish asymptotically
to both the future and to the past. $\Sigma_A=\Sigma_C=0$
defines a circle of fix points, the projected Kasner circle:
${\rm KP}^\ocircle$, see Figure~\ref{Ksectors}. It is described
by $\Sigma_+=\hat{\Sigma}_+,\Sigma_B=\hat{\Sigma}_-$, where the
constants $\hat{\Sigma}_+$, $\hat{\Sigma}_-$ satisfy
$\hat{\Sigma}_+^2 + \hat{\Sigma}_-^2=1$.

From this we conclude that the the $\alpha$-limits for all
solutions with $\Sigma_A\Sigma_C\neq 0$ on ${\cal KP}$ resides
on the local source of ${\rm KP}^\ocircle$, yielding a segment
on ${\rm KP}^\ocircle$ characterized by
$-1\leq\Sigma_+=\hat{\Sigma}_+\leq -\frac{1}{2},\, 0\leq
\hat{\Sigma}_-\leq\frac{\sqrt{3}}{2}$, i.e., the segment
consists of sector $(213)$ together with the fix points ${\rm
Q}_2$ and ${\rm T}_3$ on ${\rm KP}^\ocircle$. The
$\omega$-limit resides on the local sink, which consists of
segment $(312)$ together with the fix points ${\rm Q}_3$ and
${\rm T}_2$, see Figure~\ref{Kasnersourcesink}

The $\alpha$-limits for solutions on ${\cal K}$ are determined
by the $\alpha$-limits on ${\cal KP}$ which determine the
asymptotic limits for $v_{(i)}$. The equation for
$|v_{(i)}|\in[0,1]$ on ${\rm KP}^\ocircle$ is given by:
$|v_{(i)}|^\prime = (G^{(i)}_-)^{-1} (1-v_{(i)}^2)(3w - 1 +
2\hat{\Sigma}_+) \,|v_{(i)}|$. It follows that the
$\alpha$-limits for all orbits on ${\cal K}$ on the general
geometric set with $\Sigma_A\Sigma_C\neq 0$ resides on the
global past attractor ${\cal A}_{\{**\}}$, where the subscript
denotes the range of values of $w$, given by
\begin{subequations}\label{Kattr}
\begin{align}
{\cal A}_{\{w<\sfrac{2}{3}\}} =
& \{{\rm K}^\ocircle_{11}: \hat{\Sigma}_+\in \left[-1,-\sfrac{1}{2}\right]\}\:, \label{Kattra}\\
{\cal A}_{\{w=\sfrac{2}{3}\}} =
& \{{\rm K}^\ocircle_{11}: \hat{\Sigma}_+\in \left[-1,-\sfrac{1}{2}\right)\}\cup
\{{\rm KS}^+_{v_{(1)}v_{(2)}}: \hat{\Sigma}_+=-\sfrac{1}{2}\}\:,\\
{\cal A}_{\{\sfrac{2}{3}<w\}} =
& \{{\rm K}^\ocircle_{11}: \hat{\Sigma}_+\in \left[-1,-\sfrac{1}{2}(3w-1)\right)\}\cup
\{{\rm KS}^+_{v_{(1)}v_{(2)}}: \hat{\Sigma}_+=-\sfrac{1}{2}(3w-1)\}\cup
\nonumber\\
& \{{\rm K}^\ocircle_{00}: \hat{\Sigma}_+\in\left(-\sfrac{1}{2}(3w-1),-\sfrac{1}{2}\right]\}\:.
\end{align}
\end{subequations}
The local stability analysis of the Kasner circle in the full
Bianchi type I state space shows that the entire circle has
unstable modes and will therefore not attract any orbits
outside of the Kasner subset; the sinks on ${\cal K}$ are
saddle points in the full state space.

\end{appendix}

\end{document}